\documentclass[10pt, conference]{IEEEtran}
\IEEEoverridecommandlockouts{}

\usepackage{amsmath,amssymb,amsfonts}
\usepackage{algorithmic}
\usepackage{graphicx}
\usepackage{textcomp}
\usepackage{xcolor}

\usepackage{qcircuit}
\usepackage{adjustbox}
\usepackage[ruled,linesnumbered,vlined]{algorithm2e}
\usepackage{braket}
\usepackage{threeparttable}

\usepackage{soul}
\usepackage{balance}

\usepackage[colorlinks,linkcolor=blue,anchorcolor=blue,citecolor=blue]{hyperref}

\SetNlSty{textbf}{}{}

\newcommand{\ZZ}{\ensuremath{\mathrm{ZZ}} }

\newcommand{\surgeq}{\textsc{SurgeQ}}
\newcommand{\qiskit}{\textsc{Qiskit}}
\newcommand{\tket}{\textsc{TKet}}
\newcommand{\pyscf}{\textsc{PySCF}}

\def\BibTeX{{\rm B\kern-.05em{\sc i\kern-.025em b}\kern-.08em
    T\kern-.1667em\lower.7ex\hbox{E}\kern-.125emX}}

\SetAlFnt{\small}
    
\begin{document}

\title{\surgeq: A Hybrid Framework for Ultra-Fast Quantum Processor Design and Crosstalk-Aware Circuit Execution
}

 \author{
     \IEEEauthorblockN{
     Xinxuan Chen \IEEEauthorrefmark{1}\textsuperscript{1}, Hongxiang Zhu\IEEEauthorrefmark{1}\textsuperscript{1}\thanks{\textsuperscript{1}Co-first authors.}, 
     Zhaohui Yang\IEEEauthorrefmark{2}, 
     Zhaofeng Su\IEEEauthorrefmark{1}\IEEEauthorrefmark{3},
     Jianxin Chen\IEEEauthorrefmark{4},
     Feng Wu\IEEEauthorrefmark{5}\textsuperscript{2},
     Huihai Zhao\IEEEauthorrefmark{5}\textsuperscript{2}
     \thanks{\textsuperscript{2}Corresponding authors. \href{mailto:zhaohuihai@iqubit.org}{zhaohuihai@iqubit.org}, \href{mailto: wufeng@iqubit.org}{wufeng@iqubit.org}.}
     }
    
     \IEEEauthorblockA{\textit{\IEEEauthorrefmark{1}School of Computer Science and Technology, University of Science and Technology of China, Hefei 230026, China}}
     \IEEEauthorblockA{\textit{\IEEEauthorrefmark{2}Department of Electronic and Computer Engineering, The Hong Kong University of Science and Technology, Hong Kong}}
     
     \IEEEauthorblockA{\textit{\IEEEauthorrefmark{3}Research Institute of Quantum Technology, The Hong Kong Polytechnic University, Hong Kong}}
     \IEEEauthorblockA{\textit{\IEEEauthorrefmark{4}Department of Computer Science and Technology, Tsinghua University, Beijing 100084, China}}
     \IEEEauthorblockA{\textit{\IEEEauthorrefmark{5}Zhongguancun Laboratory, Beijing 100194, China}}
    \thanks{Supported by Zhongguancun Laboratory, and sponsored by Quantum Science and Technology-National Science and Technology Major Project (Grant No. 2021ZD0302901)}
}

\maketitle

\begin{abstract}

Executing quantum circuits on superconducting platforms requires balancing the trade-off between gate errors and crosstalk. To address this, we introduce \surgeq, a hardware-software co-design strategy consisting of a design phase and an execution phase, to achieve accelerated circuit execution and improve overall program fidelity. \surgeq\ employs coupling-strengthened, faster two-qubit gates while mitigating their increased crosstalk through a tailored scheduling strategy.
With detailed consideration of composite noise models, we establish a systematic evaluation pipeline to identify the optimal coupling strength. Evaluations on a comprehensive suite of real-world benchmarks show that \surgeq\ generally achieves higher fidelity than up-to-date baselines, and remains effective in combating exponential fidelity decay, achieving up to a million-fold improvement in large-scale circuits.

\end{abstract}

\section{Introduction}

To achieve high-fidelity execution of quantum applications, a quantum processor must adhere to specific design principles. In particular, errors related to the two-qubit gates have been identified as the primary concern in practical large-scale quantum computing~\cite{chen2021exponential,mi2021information,acharya2023suppressing}.
Due to their sensitivity to noise-induced decoherence, quantum processors benefit from shorter two-qubit gate times through strengthened couplings between adjacent qubits.
On the other hand, executing quantum circuits swiftly by using parallel operations is crucial to mitigate decoherence effects. However, fast operations and parallelism are sometimes mutually exclusive.
Stronger qubit coupling exacerbates stray interactions, leading to severe crosstalk errors between simultaneous two-qubit gates and heightening the need for scheduling strategies to mitigate the spatio-temporal crosstalk.
Balancing these two kinds of errors calls for a deliberate selection of coupling strength and gate time.

\begin{figure}[tbp]
    \centering
    \includegraphics[width=1\linewidth]{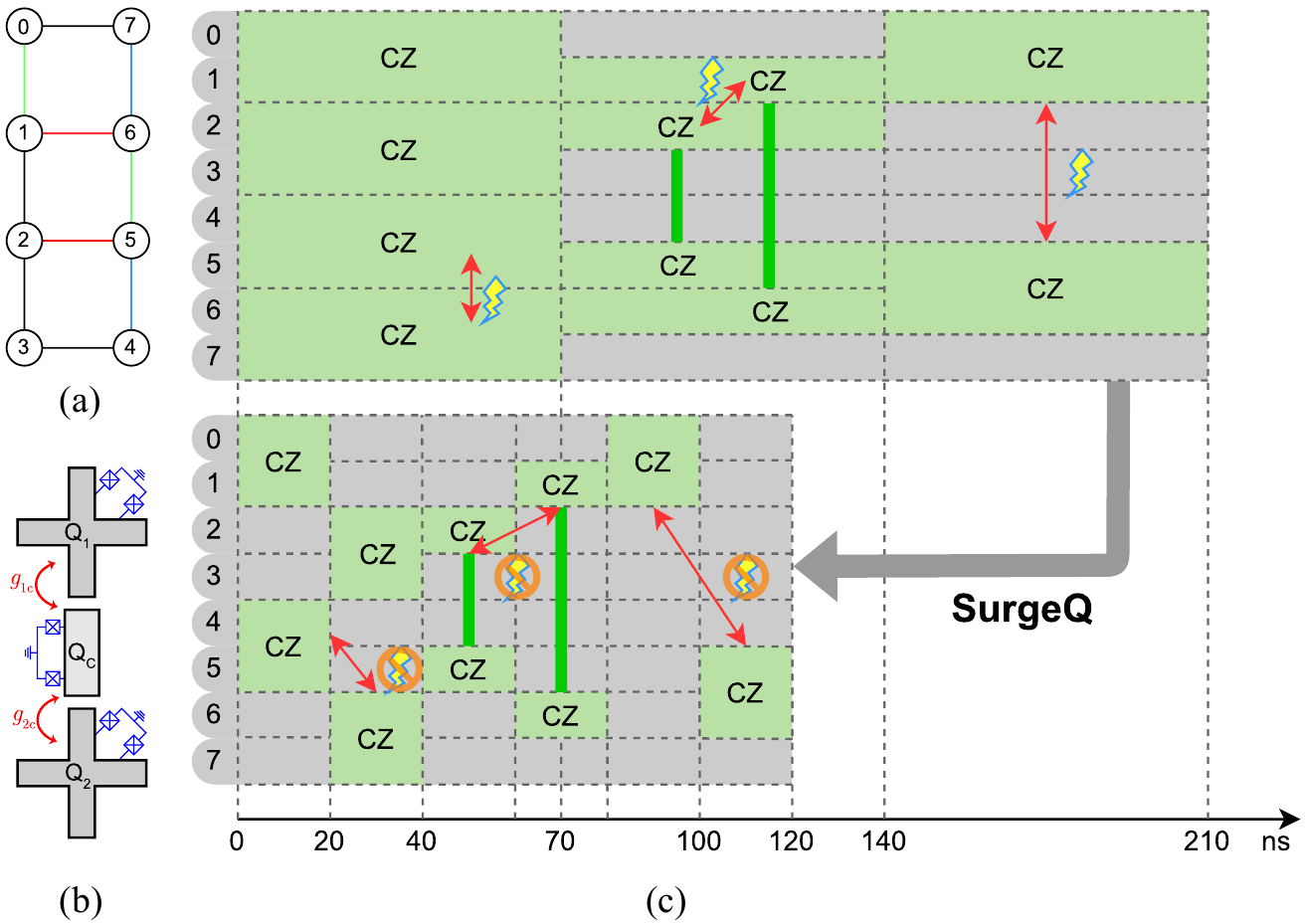}
    \caption{Overview of \surgeq's strategies on circuit execution. (a) Qubit connectivity graph for the example circuit. (b) Illustration of tunable couplers with adjustable coupling strengths. (c) Execution schedules before (top) and after (bottom) applying \surgeq. The initial schedule contains multiple simultaneous CZ gates on adjacent qubits, leading to crosstalk events (highlighted with red arrows and yellow lightning markers, corresponding to colored coupling paths in (a)). The \surgeq-optimized schedule jointly applies gate-level scheduling and coupling-strength tuning to resolve all crosstalk conflicts and significantly reduce execution time.}\label{fig:SurgeQ}
\end{figure}

Crosstalk, as highlighted above, becomes particularly severe under stronger couplings and parallel two-qubit operations, and remains one of the most significant challenges in executing large-scale circuits.
To address this, several strategies have been introduced. For example, frequency allocation and adjustment techniques are employed to mitigate the detrimental effects of stray interactions caused by frequency crowding~\cite{ding2020systematic, zhao2022quantum}. Other solutions target different stages of the compilation process.~\cite{xie2022suppressing} introduces a collaborative approach that integrates pulse optimization with scheduling, while~\cite{lu2024camelphysicallyinspiredcrosstalkaware} incorporates crosstalk-aware mapping strategies alongside gate scheduling. In contrast,~\cite{murali2020software} explores the trade-off between crosstalk mitigation and decoherence error reduction, employing optimization algorithms to minimize the impact of various errors within the circuit. 

Crosstalk is not unique to quantum computing; its classical counterparts have been extensively studied in areas like integrated circuits, wireless communication, and signal processing~\cite{Duan2004,Sinha_Kar_Bhattacharjee_2010, Daraban2012}. 
Inspired by the idea of crosstalk cancellation used to accelerate on-chip buses~\cite{Duan2004}, we propose and develop \surgeq---a hardware–software co-design framework implemented in two phases: a design stage that strengthens couplings to realize faster two-qubit gates, and an execution stage employing tailored scheduling to suppress the amplified crosstalk resulting from two-qubit gate acceleration. 
Unlike previous crosstalk-aware scheduling approaches relying on software-only mitigation~\cite{murali2020software}, frequency allocation~\cite{ding2020systematic}, or pulse-level co-optimization without design-stage considerations~\cite{xie2022suppressing}, our scheduler is explicitly tailored to co-design with hardware-level gate acceleration. \surgeq\ first leverages stronger coupling to realize faster two-qubit gates, and then applies scheduling to suppress the amplified crosstalk, a capability not addressed in prior works~\cite{ding2020systematic, xie2022suppressing, lu2024camelphysicallyinspiredcrosstalkaware, murali2020software}.
Fig.~\ref{fig:SurgeQ} illustrates the strategies employed by \surgeq\ to improve circuit execution. It contrasts an initial gate execution scenario with the result of applying \surgeq, which jointly applies gate-level scheduling and coupling-strength tuning to resolve crosstalk-inducing conflicts and reduce execution time. While the specific circuit used is simplified, the visual comparison conveys the key intuition behind \surgeq. The process begins with increasing the qubit coupling strength to enable faster gates, which is critical for improving overall execution performance. However, stronger coupling significantly amplifies crosstalk, making naive speed-up detrimental to circuit fidelity. To address this, \surgeq\ tightly integrates gate scheduling with coupling-strength calibration, jointly restructuring the circuit to eliminate crosstalk while preserving the benefits of faster gates. The result is a synergistic effect that surpasses what either technique can achieve in isolation.

Having established \surgeq\ as our co-design framework, we now highlight the concrete contributions as follows:
\begin{enumerate}
    \item [1.] We propose \surgeq, a co-designed framework that significantly accelerates circuit execution by tuning coupling strength to maximize two-qubit gate speed, while effectively mitigating the exacerbated crosstalk effects through circuit optimization and scheduling.
    \item [2.] We introduce a systematic evaluation pipeline that fully accounts for the dominant error sources to identify the feasible optimal coupling strength, and show that omitting any stage of our strategy markedly degrades overall performance.
    \item [3.] We further validate \surgeq's effectiveness and versatility through a comprehensive evaluation on a diverse suite of real-world–inspired quantum circuits. Compared with other baselines, \surgeq\ achieves higher fidelity in most cases and remains effective in combating exponential fidelity decay, reaching up to a million-fold improvements on larger-scale circuits.
\end{enumerate}

\section{Background}\label{sec:background}

Two-qubit gates present a greater challenge than single-qubit gates for quantum processors' performance~\cite{zajac2018resonantly,wright2019benchmarking,chen2021exponential,mi2021information,acharya2023suppressing}. Issues such as stray interactions due to the weak anharmonicity of transmons~\cite{krantz2019quantum,blais2021circuit} and increased frequency crowding as systems scale up~\cite{arute2019quantum,han2020error} have been partially mitigated by tunable couplers. Such a design facilitates high-fidelity two-qubit gates in scalable processors.

Tunable couplers connect two qubits through electromagnetic interactions~\cite{yan2018tunable}. Specifically, the strength of this interaction can be adjusted by modifying the inductance and capacitance~\cite{chen2014qubit}. The Hamiltonian describing two transmon qubits connected by a transmon tunable coupler, as illustrated in Fig.~\ref{fig:SurgeQ}(b), can be approximately given by (hereafter $\hbar=1$)
$H = \sum_{i=1,2,c}(\omega_i a_i^+a_i + \frac{\alpha_i}{2}a_i^+a_i^+a_i a_i) +\sum_{i=1,2}g_{ic}(a_i+a_i^+)(a_c+a_c^+)$, where the subscript $i\in\{1,2,c\}$ labels transmon $Q_i$. Here $a_i^+$ and $a_i$ are the corresponding creation and annihilation operators, $\omega_i$ is the qubit frequency, $\alpha_i$ is the anharmonicity, and $g_{ic}$ denotes the coupling strength between two qubits $Q_i$ and $Q_c$.
When the coupler is in its ground state, the \ZZ interaction between $Q_1$ and $Q_2$ is defined as
$\zeta_{\ZZ}\equiv (E_{101}-E_{100})-(E_{001}-E_{000})$, where $E_{jkl}$ denotes the eigenenergy associated with the eigenstate $|jkl\rangle$ in the eigenbasis $\ket{Q_1,C,Q_2}$. The tunable coupler can be biased to tune up the coupling strength between $\ket{101}$ and $\ket{002}$, shifting $E_{101}$ and activating the \ZZ interaction, which scales as $g_{1c}g_{2c}$~\cite{yan2018tunable,sung2021realization}. By accumulating this \ZZ interaction over time, we can natively implement any gate from the $\mathrm{CPhase}(\theta)$ family~\cite{krantz2019quantum}. Usually the $\mathrm{CZ}=\mathrm{CPhase}(\pi/2)$ gate is calibrated on hardware to serve as the dominant two-qubit basis gate. Faster $\mathrm{CZ}$ operation can be realized via a larger \ZZ interaction within the coupling strength bound.

However, there are non-negligible crosstalk errors among superconducting qubits arising from spatial and temporal correlations of quantum operations, especially for qubits with larger interactions~\cite{sarovar2020detecting}. In this work, we focus on the spatially non-local effects of gates, particularly the adjacent parallel two-qubit gates. In tunable coupler architectures, adjacent parallel two-qubit gates increase the likelihood of higher energy levels of couplers and qubits approaching each other, leading to significantly larger errors, compared to isolated two-qubit gates~\cite{zhao2022quantum}. Such interaction creates unwanted \ZZ terms across multiple qubits and couplers.

\section{Proposed Framework}\label{proposed framework}

\subsection{Overview}

\begin{figure}[tbp]
    \centering
    \includegraphics[width=0.6\linewidth]{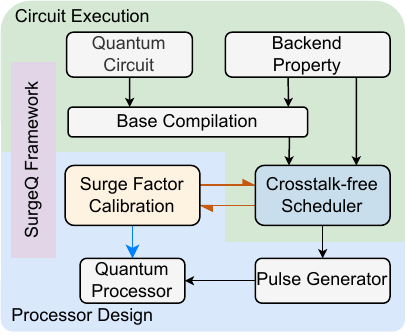}
    \caption{Overview of the \surgeq\  framework. In addition to the standard workflow of quantum processor design and circuit execution, the Surge Factor Calibration Unit is introduced to evaluate performance gains achieved through highly optimized circuit execution that avoids crosstalk penalties, as indicated by the bi-directional red arrows. The optimal Surge factor guides the selection of appropriate coupling strengths, as shown by the blue arrow.}\label{fig:Workflow}
\end{figure}

Figure~\ref{fig:Workflow} depicts the workflow of the proposed \surgeq\ framework. To fairly evaluate the trade-offs of accelerating quantum operations, we compare all scenarios using highly optimized quantum circuits. 
While circuit optimization is an active area of research, we emphasize simplicity and leverage third-party SDKs such as \qiskit\ to highlight our core contributions.
A quantum circuit intended for execution on a chip is processed through a compilation workflow borrowed from third-party SDKs, referred to as {\em base compilation}. This workflow incorporates the backend properties of the chip, such as qubit connectivity topology, and outputs a compatible optimized circuit. We then introduce a {\em Crosstalk-Free Scheduler} that prevents parallel operations on adjacent qubit pairs, thereby eliminating nearest-neighbor \ZZ crosstalk at the cost of increasing circuit depth. Unlike \cite{murali2020software}, our scheduler does not rely on physical characterization data and can reduce the depth overhead of the serialization process.

We introduce a {\em Surge factor}, $s$, which indicates the amplification of the coupling strength during the processor design phase to balance errors from different sources. Specifically, $g_{1c} $ and $g_{2c}$ are scaled to $s$ times their respective reference values. It is optimized in the {\em Surge Factor Calibration} module to determine the optimal $s$ that maximizes circuit fidelity.

In the following subsections, we will provide a detailed discussion of both modules.

\subsection{Surge Factor Calibration}\label{subsec:error}

The Surge Factor Calibration module begins by extracting hardware parameters, such as qubit connectivity, gate durations, decoherence time and crosstalk, from specifications or standard characterization experiments on a testing quantum processor. Using these parameters, this module estimates errors of this processor, which is provided to a scheduler that performs crosstalk-aware gate scheduling, simulates circuit execution to compute the fidelity across a range of surge factors. Then this module identifies an optimal surge factor to be applied in a new processor design intended to be fabricated as the final hardware to execute the circuit. 

The physical implementation of the design depends on the processor architecture. For fixed coupling architectures, tuning the coupling between qubits is achieved by adjusting the geometry (i.e., size and shape) of qubits. In this work, we take tunable coupling architecture as an example, of which the geometry of both qubits and couplers requires modification.

To evaluate the effect of the Surge Factor $s$ on quantum gate performance, this module employs a comprehensive error model. In this model, the relationships between different error sources and the Surge Factor $s$ are listed below.

\begin{algorithm}[tp]
    \SetKwProg{subproc}{Procedure}{}{}
    \DontPrintSemicolon
    \caption{Crosstalk-Free Scheduling}
    \label{alg:free_crosstalk_schedule}
    \KwIn{A quantum circuit with timing information}
    \KwOut{A scheduled DAG without \ZZ crosstalk}
    $\mathcal{D}=\{V_D,E_D\}$ $\leftarrow$ translate input circuit into DAG \;
    $\mathcal{T}=\{V_T,E_T\}$ $\leftarrow$ initialize an empty DAG \;
    \While{$\mathcal{D} \ne \varnothing$}{
        $V_0$ $\leftarrow$ head nodes of $\mathcal{D}$ \;
        $S(C)$ $\leftarrow$ nodes of single-(two-)qubit gates in $V_0$ \;\label{line:choose2q}
        $M$ $\leftarrow$ \textsc{MISGates}($C$) \;\label{line:MISGates}
        $R=C-M$ \;
        $R^{\prime}$ $\leftarrow$ \textsc{PatternSubstitute}($R$, $\mathcal{D}$) \;\label{line:pattern}
        Layer $\leftarrow$ $S \cup M \cup R^{\prime}$ \;
        $\mathcal{T}$.add(Layer) \;\label{line:transfer_begin}
        $\mathcal{D}$.delete(Layer) \;\label{line:transfer_end}
    }
    \Return{$\mathcal{T}$} \;
    \subproc{\textnormal{\textsc{MISGates}(Node set $C$)}\label{line:MISGates_begin}}{ 
        $\mathcal{G}$ $\leftarrow$ init an empty undirected graph $\mathcal{G}(V,E)$ \;
        $\mathcal{G}.V \leftarrow C$ \;
        \ForEach{$\{v_1, v_2\} \subset \mathcal{G}.V$}{
            \If{$v_1$ and $v_2$ have nearest \ZZ crosstalk\;}{$\mathcal{G}.E$.add($(v_1,v_2)$)
            }
        }
        $M$ $\leftarrow$ nodes that constitute the MIS of $\mathcal{G}$ \;
        \Return{$M$} \;\label{line:MISGates_end}
    }
    \subproc{\textnormal{\textsc{PatternSubstitute}(Nodes $V$, DAG $\mathcal{D}$)}\label{line:PatternSubtstitute_begin}}{
        \ForEach{$v \in V$}{
            \If{$v$ and its successors in $\mathcal{D}$ match a pattern\;\label{line:recognition}}{
                Substitute the pattern in $\mathcal{D}$ \;\label{line:substitution}
                $s_1,s_2$ $\leftarrow$ the new head nodes \;
                $R^{\prime}$.add($s_1,s_2$) \;
            }
        }
        \Return{$R^{\prime}$} \;\label{line:PatternSubtstitute_end}
    }
\end{algorithm}
\begin{enumerate}
    \item \textbf{Two-Qubit Decoherence Error:} The two-qubit gate error can be attributed to multiple sources~\cite{xu2020high}. First, we assume the two-qubit gate time $t_g$ is inversely proportional to the qubit-qubit \ZZ interaction strength. Based on the discussion in the previous section, this implies $s^2 \propto 1/t_g$. We decompose the decoherence error into two components: $E_1$ and $E_{\phi}$. The relaxation error $E_1$, arising from qubit relaxation ($T_1$), follows the relation $E_1=\frac{1}{3}t_g/T_1$, which is proportional to the gate time, yielding $E_1 \propto t_g \propto s^{-2}$. For pure dephasing, assumed to be dominated by $1/f$ flux noise, the error is given by $E_{\phi}=\frac{1}{3}t_g^2/T_{\phi}^2$, proportional to the square of the gate time, leading to $E_{\phi} \propto t_g^2 \propto s^{-4}$. 

    \item \textbf{Two-Qubit Coherent Error:} The dominant coherent error in the two-qubit gate, the non-adiabatic error $E_A$, is determined by the adiabatic condition factor and the pulse shape~\cite{xu2020high}. However, there is no simple relationship between $s$ and $E_A$. Considering that varying $s$ affects $E_A$ through two opposing mechanisms---larger energy level splitting from stronger coupling and faster energy changes from shorter gate times---which can partially offset each other, we assume $E_A$ remains constant for different $s$ in this work.

    \item \textbf{Single-Qubit Error:} On the other hand, the coupling strength does not affect single-qubit gates. Thus, the decoherence errors $E_1$, $E_{\phi}$, and the coherent error $E_c$ of single-qubit gates remain constant irrespective of $s$.

    \item \textbf{Crosstalk Effects on Single-Qubit Errors:} Crosstalk errors arise when adjacent qubits execute single-qubit or two-qubit gates simultaneously~\cite{zhao2022quantum}. For single-qubit gates, crosstalk manifests as cross-drive on nearby qubits, leading to an error $E_{1Q}$. This error scales with the square of the cross-drive strength, which is proportional to the qubit-qubit coupling strength, resulting in $E_{1Q} \propto s^4$. 

    \item \textbf{Crosstalk Effects on Two-Qubit Errors:} For parallel two-qubit gates, the error $E_{2Q}$ originates from stray \ZZ interactions involving higher energy levels of four qubits, such as $\ket{0002}$ and $\ket{1000}$. These interactions scale as $s^6$, while the gate time scales as $s^{-2}$, leading to $E_{2Q} \propto s^4$.
\end{enumerate}

\subsection{Crosstalk-Free Scheduler}

Our scheduler staggers adjacent two-qubit gates while minimizing additional circuit depth. We model circuits as Directed Acyclic Graphs (DAGs), where nodes represent gates and edges represent dependencies. We distinguish the layer structure by inserting barrier nodes between consecutive layers, thereby dividing the DAG into layers and enabling sequential scheduling.

\begin{figure}
    \centering
    \includegraphics[width=1\linewidth]{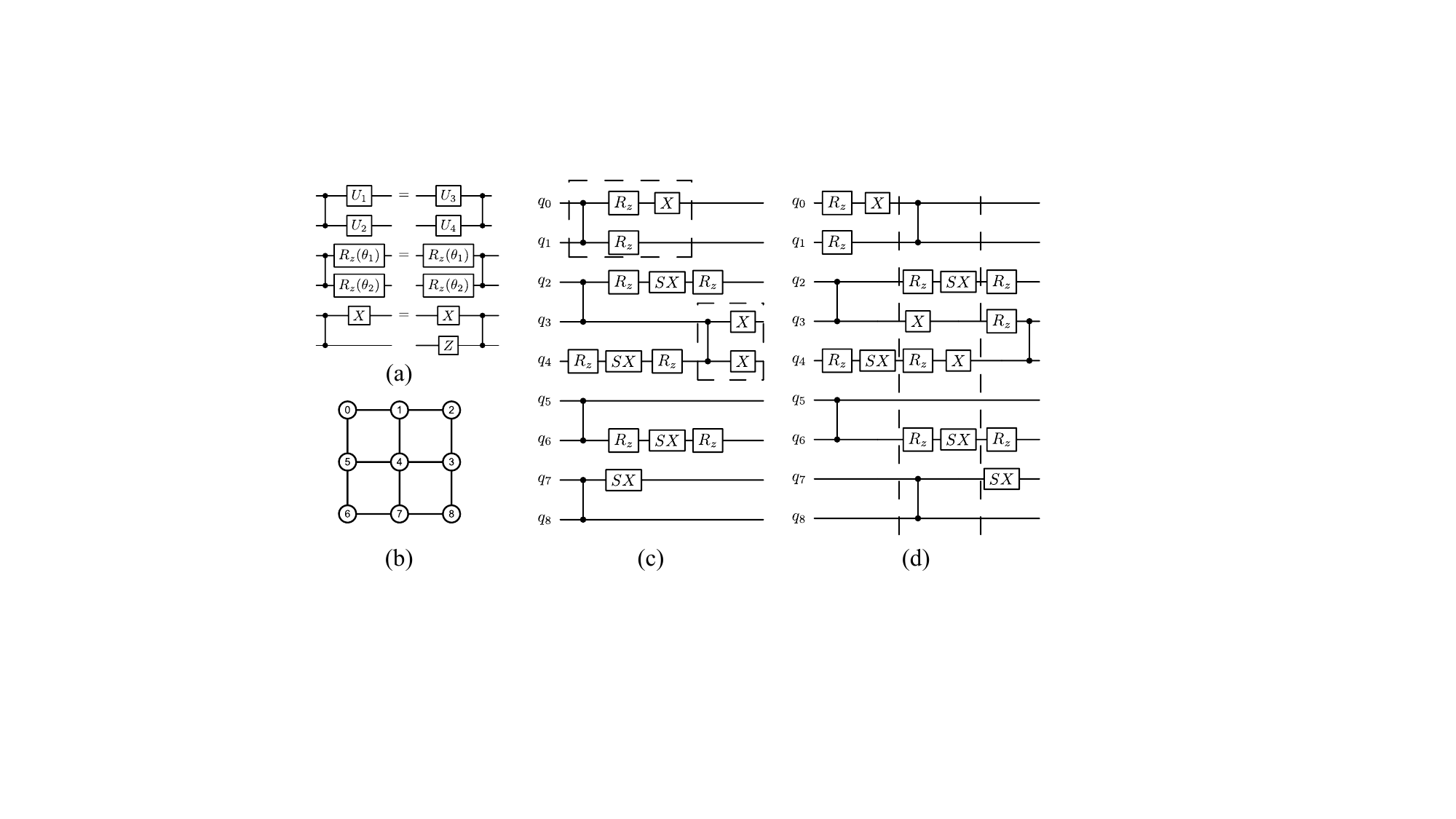}
    \caption{(a) Pattern substitution demonstration. The topmost pattern represents the desired form of pattern substitution. The rest patterns illustrate the current set of pattern substitutions available in our library. (b) Example topological layout of the qubits for (c)(d). The nodes in the topological layout represent qubits while edges denote the coupling between qubits. (c)(d) An example of a scheduling algorithm for the quantum circuit (c) with result (d). We omit the rotation angle $\theta$ around the $Z$ axis of $R_z(\theta)$ gate, as the pattern substitution and merging of $R_z(\theta)$ gates involving changes in $\theta$ of certain $R_z(\theta)$ are irrelevant to our approach. 
    The duration of the whole layer is decided by the longest gate inside, with the timing inconsistency compensated by idle time in each layer. The enclosed parts in (c) are identified as structures that can undergo pattern substitution. In the first loop of the algorithm, the two-qubit gate set $\{q_2\text{-}q_3,q_5\text{-}q_6\}$ is the MIS found, while a pattern containing $q_0$-$q_1$, along with their subsequent single-qubit gates, has been identified. The same line of thought is also applied in the following loop.
    }
    \label{fig:pattern}
\end{figure}

Algorithm~\ref{alg:free_crosstalk_schedule} converts the input circuit into a DAG $D$ and iteratively extracts head nodes $V_0$. Among two-qubit gates $C \subseteq V_0$, we apply a Maximum Independent Set (MIS) algorithm to select the largest subset $M$ free of nearest-neighbor \ZZ crosstalk. Remaining gates $R$ are checked for predefined substitution patterns, producing modified nodes $R^{\prime}$. Together with single-qubit gates $S$, these ($M$, $R^{\prime}$) form a scheduled layer, appended to the target DAG $T$. The process repeats until $D$ is empty, with all $R_z(\theta)$ rotations excluded from the layer count (e.g.,\ ``$-R_z(\theta)-SX-$'' is treated as one layer). The final DAG fully eliminates nearest-neighbor \ZZ crosstalk. An example is illustrated in Fig.~\ref{fig:pattern}(b)(c)(d).

Within the MIS algorithm, two-qubit gates are mapped to an undirected graph where edges indicate potential crosstalk. The MIS gives the maximum parallel set, though deferred gates increase idle time and circuit depth. To mitigate this, we apply pattern substitution: if some deferred gates and their successors match a predefined template, a substitution is applied. This substitution reduces the overhead introduced by MIS, and as the pattern library expands, the scheduler can further minimize depth growth. The desired form of pattern substitution is illustrated at the top of Fig.~\ref{fig:pattern}(a), where the rest illustrates the current set of pattern substitutions available in our library.

\section{Evaluation}

\subsection{Experimental Setup}

\subsubsection{Digital error model}
Digital quantum computing relies on the assumption that errors are discrete, localized, and independent both temporally and spatially. This assumption was experimentally validated in~\cite{arute2019quantum}, where a digital error model was also proposed to characterize the expected behavior of large-scale quantum circuits.

Error models for individual gates are thoroughly discussed in Subsec.~\ref{subsec:error}. The digital error model describes the expected behavior of a circuit composed of these individual gates, focusing on accuracy or fidelity within the context of quantum computing. Under the assumption of the digital error model, the circuit fidelity is calculated as the product of the fidelities of individual gates.

We note that during the execution of a two-qubit gate, one qubit is tuned away from its idle point, resulting in a significant reduction in its effective $T_1$ and $T_{\phi}$ due to stronger interaction with the coupler and increased flux noise~\cite{xu2020high}. In the equations below, the superscript ``eff'' denotes the qubit in this condition, while ``idle'' is used to represent the qubit in its idle state.

The fidelity of a single-qubit gate is given by $F_{\mathrm{1Q}} = 1 - E_c - E_1^{\mathrm{idle}} - E_{\phi}^{\mathrm{idle}} - N_1 E_{\mathrm{1Q}}$, where $N_1$ is the number of single-qubit gates adjacent to this single-qubit gate. Similarly, the fidelity of a two-qubit gate is given by $F_{\mathrm{2Q}} = 1 - E_1^{\mathrm{eff}} - E_1^{\mathrm{idle}} - E_{\phi}^{\mathrm{eff}} - E_{\phi}^{\mathrm{idle}} - E_A - N_1 E_{\mathrm{1Q}} - N_2 E_{\mathrm{2Q}}$. 
Here, $N_1$ and $N_2$ count the single- and two-qubit gates adjacent to this two-qubit gate. When the qubit is idle, the fidelity decreases as $F_{\mathrm{idle}}=1 - E_1^{\mathrm{idle}} - E_{\phi}^{\mathrm{idle}}$. 
Finally, the circuit fidelity is calculated by $F = \prod_{\{\mathrm{1Q}\phantom{.}\mathrm{gate}\}}F_{\mathrm{1Q}} 
    \prod_{\{\mathrm{2Q}\phantom{.}\mathrm{gate}\}}F_{\mathrm{2Q}} 
    \prod_{\{\mathrm{idle}\phantom{.}\mathrm{time}\}}F_{\mathrm{idle}}$.

\begin{table}[tbp]
    \centering
    \renewcommand\arraystretch{1.1}
    \caption{Physical Parameters Realizable with Current Hardware.}\label{tab:Physical Parameters}
    \begin{tabular}{|c|c|c|c|}
        \hline
        $T_{1}^{\mathrm{eff}}$ & $T_{\phi}^{\mathrm{eff}}$ & $T_{1}^{\mathrm{idle}}$ & $T_{\phi}^{\mathrm{idle}}$ \\
        \hline
        40000 ns & 1100 ns & 50000 ns & 40000 ns \\
        \hline
        $E_c$ & $E_{A}$ & $E_{\mathrm{1Q}}$ & $E_{\mathrm{2Q}}$\\
        \hline
         $3.25\times10^{-6}$ & $1.2\times10^{-3}$ & $2.13\times10^{-5}$ & $1.75\times10^{-4}$ \\ 
        \hline
    \end{tabular}
    
\end{table}

\begin{figure*}[htbp]
    \centering
    \includegraphics[width=0.97\linewidth]{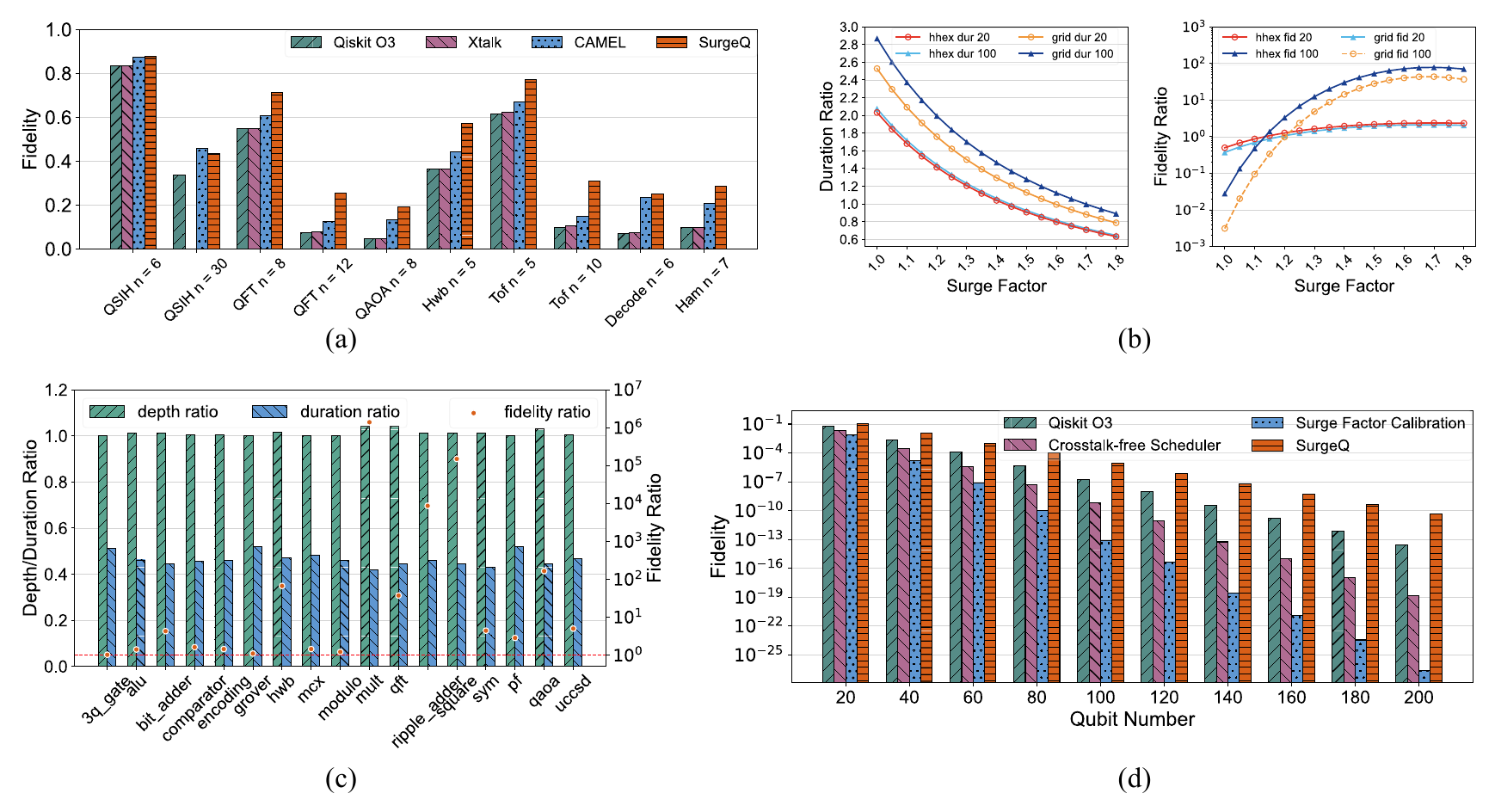}
    \caption{(a) Comparison of fidelity across several representative circuit families under different baselines. All circuits are from the dataset described in~\ref{Chap:Benchmarks}, and n denotes the number of qubits, with QSIH circuits fixed to 2 Trotter steps. Due to XtalkSched's exponential-time complexity and the resulting prohibitive compilation times, results for QSIH with n=30 are unavailable. Overall, SurgeQ achieves higher fidelity than the alternatives in most cases. (b) The Surge Factor Calibration for QSIH circuits. All the data are the ratio compared with the results obtained by using only the base compilation---\qiskit\  O3. A lower duration ratio and a higher fidelity ratio are indicative of superior performance. Legends such as hhex/grid, dur/fid, and $20$/$100$ indicate the corresponding data points' attributes: topology (heavy hexagon or grid), metric (duration ratio or fidelity ratio), and circuit size ($20$-qubit or $100$-qubit). (c) Benchmarking results of \surgeq\  across a comprehensive suite of quantum circuits derived from real-world quantum applications, evaluated in terms of depth, duration, and fidelity ratios, with the Surge factor set to $s=1.8$. (d) Ablation experiment of \surgeq\  for QSIH execution on grid backend with $s=1.7$}
    \label{fig:evaluation}
\end{figure*}
\subsubsection{Setup}
Our compilation experiments are conducted on an Intel Core i5-14500 processor running Linux 6.8.0. We implement our approach in Python 3.9, utilizing \qiskit\  1.0.2~\cite{qiskit2024} and NetworkX 3.2.1~\cite{hagberg2020networkx}. The base compilation used in our study is \qiskit\  optimization level 3 (-O3), which represents the highest optimization level for the original circuit. 

\subsubsection{Quantum Backend}
We compile quantum circuits into a basic gate set $\{I, X, SX, R_Z(\theta), CZ\}$, where $SX=\sqrt{X}$, $R_Z(\theta)=e^{-i\theta Z/2}$. Referring to~\cite{zhao2022quantum}, we set the time of single-qubit gates $S$ and $SX$ to $20$ ns, and initialize the CZ gate with a duration of 70 ns without adjusting the coupling strength. According to the Virtual Z technology~\cite{mckay2017efficient}, we set the time of $R_Z(\theta)$ as $0$ for all $\theta$ available. The backends we are using include a 127-qubit heavy-hex (hhex in brief) structure and a 225-qubit grid structure. 

\subsubsection{Benchmarks}\label{Chap:Benchmarks}
To conduct thorough benchmarking, we carefully select diverse categories of quantum programs, spanning 17 categories and a total of $127$ benchmarks. Most programs are drawn from established benchmark suites~\cite{wille2008revlib,sivarajah2020t,zulehner2018efficient}, with the exception of the {\em ripple\_adder} and {\em qaoa} categories, which are generated by us. All the categories include {\em 3q\_gate}, {\em alu} (Arithmetic Logic Unit), {\em bit\_adder}, {\em comparator}, {\em encoding} (encoding function), {\em grover}, {\em hwb} (Hidden Weighted Bit function), {\em mcx} (Multi-controlled-X gates), {\em modulo}, {\em mult} (Galois Field Multiplier), {\em qft} (Quantum Fourier Transform), {\em ripple\_adder} (Ripple Carry Adder), {\em square} (Square Root function), {\em sym} (Symmetric function), {\em pf} (1D Heisenberg Product Formula), {\em qaoa}, and {\em uccsd} (UCCSD programs generated using \qiskit\  and \pyscf~\cite{sun2018pyscf} from \tket 's published benchmark suite). This diverse selection provides a comprehensive evaluation of \surgeq's performance across a broad spectrum of quantum applications. We also select the second-order Trotterization of the Isotropic Heisenberg Hamiltonian evolution with open boundary conditions (referred to as QSIH) from~\cite{chowdhury2024enhancing}, due to its numerous adjacent parallel operations that may result in significant overhead for the Crosstalk-Free Scheduler.

\subsubsection{Physical Parameters}\label{Chap:Physical Parameters}
Building on studies of two-qubit gates and crosstalk in devices utilizing tunable couplers~\cite{xu2020high,zhao2022quantum,wang2022towards}, the parameters in Table~\ref{tab:Physical Parameters} reflect physical parameters achievable with current hardware and are used to calculate the fidelities of single-qubit gates, two-qubit gates, and idling. 

\subsection{Experimental Results}\label{Char:Experimental Results}
We will now systematically evaluate \surgeq\  using the experimental setup described above. Since our scheduling algorithm (Alg.~\ref{alg:free_crosstalk_schedule}) is heuristic and relies on a NetworkX approximation algorithm, we take the result with the minimum depth observed over 20 independent runs.

\subsubsection{Baselines}
We evaluate the efficacy of our framework by comparing it with several up-to-date baselines. We use \qiskit\ O3 as a crosstalk-agnostic optimizer and XtalkSched~\cite{murali2020software} and CAMEL~\cite{lu2024camelphysicallyinspiredcrosstalkaware} as crosstalk-aware baselines. For compatibility with CAMEL's scheduling scheme, we set the CZ gate time for it alone to 40 ns. As shown in Fig.~\ref{fig:evaluation}(a), across diverse circuit families, SurgeQ generally attains higher fidelity than all baselines. A notable exception occurs on 30-qubit QSIH circuits, where crosstalk is particularly severe: CAMEL's compensation pulse scheme affords a greater reduction in circuit depth, yielding fidelity slightly higher than that of SurgeQ.

\subsubsection{Feasibility} 

\surgeq\  offers both advantages and disadvantages: while accelerating operations to reduce decoherence errors, the increased circuit depth caused by the Crosstalk-Free Scheduler may partially offset these benefits, thus lessening the overall performance impact. Therefore, we evaluate its feasibility using QSIH, the highly dense circuit family mentioned before, as a worst-case test scenario.

We vary the Surge factor $s$ within $[1.0, 1.8]$ to accelerate gate operations during crosstalk-aware scheduling of QSIH circuits. Values above $1.8$ are excluded as they are not reliably attainable on current hardware and can introduce confounding effects such as pulse distortion~\cite{zhao2022quantum, sung2021realization, xu2020high, arute2019quantum}. The performance gain is defined in terms of the duration ratio and fidelity ratio, calculated as the duration and fidelity of the resulting circuits of \surgeq\ divided by those of \qiskit\ O3. The evaluation is conducted on the QSIH circuit family with $10$ Trotter steps, considering different qubit counts ($20$ and $100$) across two quantum chip topologies: heavy hexagonal (hhex) and grid.

Fig.~\ref{fig:evaluation}(b) illustrates the Surge Factor Calibration, showing $s = 1.7$ as optimal for QSIH circuits. As $s$ increases, the fidelity ratio improves in the QSIH algorithm with $20$ qubits and grows even more rapidly with $100$ qubits when $1 \leq s \leq 1.7$. However, a slight decrease occurs when $s \geq 1.7$, indicating that at this point, the increase in $s$ significantly elevates $E_{\mathrm{1Q}}$, offsetting error reductions from other sources. Notably, fidelity ratios increase faster in larger-scale circuits, outperforming \qiskit\  O3 by several orders of magnitude. For instance, when $s = 1.7$ and the qubit count is $100$, fidelity ratios approach $100$. The depth overhead of the crosstalk-free-scheduled circuit is $1.61\times$ that of the non-crosstalk-free-scheduled version. In all experiments, with qubit counts ranging from 20 to 200, this overhead consistently stays below $1.7\times$.

\subsubsection{Applicability} 
We will further evaluate the applicability of \surgeq. Given that the QSIH is chosen as the worst-case test scenario, where we still observe a significant performance boost, we anticipate that \surgeq\  will demonstrate advantages across a wide range of real-world applications.

Our theoretical analysis in Subsec.~\ref{subsec:error} reveals that only $E_{1Q}$ and $E_{2Q}$ increase exponentially with $s$, while our Crosstalk-Free Scheduler eliminates $E_{2Q}$. Consequently, crosstalk errors from single-qubit gates ($E_{1Q}$) have become a key factor limiting further efficiency improvements. However, the density of $E_{1Q}$ errors varies significantly across different circuits. It is therefore not surprising that the Surge Factor Calibration selects $s=1.7$ for the highly dense QSIH, while for most of the less dense real-world applications, it selects $s=1.8$.

Figure~\ref{fig:evaluation}(c) illustrates the benchmarking results for a comprehensive suite of representative circuits from real-world applications, where the result for each circuit type is the geometric mean of the results from circuits of the same type at different scales. 
These circuits utilize fewer parallel two-qubit gates on adjacent qubits, leading to lower depth and duration overhead for crosstalk-free scheduling.
The duration ratio is a more appropriate metric in this context. 
For representativeness, some benchmarks include extremely small-scale circuits, such as 3-qubit circuits. 
All benchmarks show consistent duration improvements. Small-scale circuits still achieve significant fidelity improvements, though less apparent on a logarithmic scale. With increased system size, fidelity improvements can reach as high as $10^6$. These results demonstrate \surgeq's exceptional performance across diverse real-world applications.

\subsubsection{Ablation Study}
We further conduct an experiment to evaluate the impact of the two key modules in \surgeq---the Surge Factor Calibration and the Crosstalk-Free Scheduler. For the Surge Factor Calibration, we amplify the coupling strength to accelerate quantum operations but execute the circuits without applying the Crosstalk-Free Scheduler, resulting in crosstalk penalties. For the Crosstalk-Free Scheduler, we keep the chip unchanged and focus solely on mitigating crosstalk effects through scheduling. Fig.~\ref{fig:evaluation}(d) clearly demonstrates that the absence of any individual component in our strategy significantly degrades the overall performance.

\subsubsection{Scalability} 
The time complexity of the crosstalk-free scheduling algorithm is $O (\frac{m^2}{n})$, with $m$ and $n$ denoting the gate count and the qubit count, respectively. 
Therefore, the core functionalities of \surgeq\  will not become a bottleneck for scaling up. The \surgeq\  framework leverages third-party compiler workflows, such as \qiskit\  at optimization level O3. 
Gate-level optimizations have a complexity ranging from $O(m)$ to $O(m^2)$, while qubit mapping is $O(n^2)$.
Thus, \surgeq\  is well-suited for scaling to systems with thousands of qubits.

\section{Summary and Outlook}

\surgeq\ presents a robust solution to the critical challenge of maintaining fidelity in large-scale quantum circuits. 
By synergizing coupling-strengthened processor design with ultra-fast two-qubit gates and crosstalk-aware circuit scheduling, our co-design framework effectively mitigates the exponential decay of program fidelity. By providing an optimized and practical design for both speed and noise, \surgeq\ paves the way for realizing more complex and impactful quantum subroutines, from fault-tolerant quantum memories to application-specific modules like the Quantum Fourier Transform, advancing the pursuit of practical quantum advantage.

\bibliographystyle{IEEEtran}

\bibliography{ref}

\end{document}